\documentclass[a4paper,twocolumn,11pt,accepted=2021-08-25]{quantumarticle}
\pdfoutput=1
\usepackage[utf8]{inputenc}
\usepackage[english]{babel}
\usepackage[T1]{fontenc}
\usepackage{amsmath}
\usepackage{amssymb}
\usepackage{graphicx}
\usepackage{hyperref}

\usepackage[backend=bibtex,style=numeric-comp,sorting=none]{biblatex}
\newcommand{\ket}[1]{\left\lvert #1 \right\rangle}
\newcommand{\imag}{\operatorname{Im}}
\newcommand{\erf}{\operatorname{erf}}
\newcommand{\erfc}{\operatorname{erfc}}

\begin{document}


\title{On the experiment-friendly formulation of quantum backflow}

\author{Maximilien Barbier}
\affiliation{Max-Planck-Institut f\"ur Physik komplexer Systeme, N\"othnitzer Stra{\ss}e 38, D-01187 Dresden, Germany}
\affiliation{Center for Nonlinear Phenomena and Complex Systems, Universit\'e Libre de Bruxelles (ULB), Code Postal 231, Campus Plaine, B-1050 Brussels, Belgium}
\orcid{0000-0003-4090-9262}

\author{Arseni Goussev}
\affiliation{School of Mathematics and Physics, University of Portsmouth, Portsmouth PO1 3HF, United Kingdom}
\orcid{0000-0001-9903-3299}

\maketitle


\begin{abstract}
In its standard formulation, quantum backflow is a classically impossible phenomenon in which a free quantum particle in a positive-momentum state exhibits a negative probability current. Recently, Miller et al.~[Quantum \textbf{5}, 379 (2021)] have put forward a new, ``experiment-friendly'' formulation of quantum backflow that aims at extending the notion of quantum backflow to situations in which the particle's state may have both positive and negative momenta. Here, we investigate how the experiment-friendly formulation of quantum backflow compares to the standard one when applied to a free particle in a positive-momentum state. We show that the two formulations are not always compatible. We further identify a parametric regime in which the two formulations appear to be in qualitative agreement with one another.
\end{abstract}




\section{Introduction}
\label{intro_sec}

Quantum backflow (QB), as originally introduced in Ref.~\cite{BM94}\footnote{The existence of quantum backflow was first pointed out by Allcock~\cite{All69} and Kijowski~\cite{Kij74}, but these were Bracken and Melloy~\cite{BM94} who carried out the first systematic analysis of the phenomenon. In particular, they were the first to discuss it for normalized states.}, refers to the classically forbidden fact that a free particle may exhibit a negative probability current at a particular space-time point even though it has, with certainty, a positive momentum. This original formulation of QB is concerned with a nonrelativistic structureless quantum particle of mass $m$ that follows a free one-dimensional motion along the $x$ axis. We will be denoting the state of the particle at time $t$ by $\ket{\psi_t}$ and its position representation by $\psi_t(x) \equiv \langle x | \psi_t \rangle$. The probability current $j_t(a)$ at a position $x=a$ is then given by
\begin{align}
j_t(a) \equiv \frac{\hbar}{m} \imag \left[ \psi_t^*(x) \frac{\partial}{\partial x} \psi_t(x) \, \right]_{x=a} \, ,
\label{j_t_def}
\end{align}
where $\imag (z)$ and $z^*$ denote the imaginary part and complex conjugate of a complex number $z$, respectively. If $\ket{\psi_t}$ is a positive-momentum state, then QB occurs whenever~\cite{BM94}
\begin{align}
j_t(a) < 0
\label{QB_standard_crit}
\end{align}
at some position $a$ and time $t$.

The phenomenon of QB has been investigated in a variety of different scenarios, and many explicit examples of backflowing states have been constructed, see e.g. Refs.~\cite{BM94,EFV05,YHH12,HGL13,Bra21}. Thus, QB has been considered for a particle moving in the presence of linear~\cite{MB98} and short-range~\cite{BCL17} potentials. It has been extended to rotational motion~\cite{Str12,PPR20,Gou20_2}, as well as to relativistic~\cite{MB98_Dirac,SC18,ALS19} and many-particle systems~\cite{Bar20}. The spatial extent of QB has been addressed in Refs.~\cite{EFV05,Ber10,BCL17}. QB is also known to bear close relation to the arrival-time problem~\cite{All69,Kij74,MLP98,MPL99,ML00,GKP06,Yea10,Hal16,HBL19,DN21} and some nonclassical aspects of the flow of probability in quantum systems~\cite{AGP16,Gou19,DT19,Gou20_1,Bra21}.

As of today, QB has never been observed experimentally. It has been argued that QB can be observed in experiments with Bose-Einstein condensates~\cite{PTM13, MP14}. Recently, an experimental realization of an optical counterpart of QB has been reported~\cite{EZB20}.

Of particular significance to the present study is Ref.~\cite{MYD21} that addresses QB for a particle, moving in a potential $V(x)$, whose state $\psi_t(x)$ contains a priori both positive \textit{and} negative momenta. In order to treat this scenario, the authors propose an alternative, ``experiment-friendly'' (EF) definition of QB, which is based on replacing the right-hand side of Eq.~\eqref{QB_standard_crit} by an integral involving the negative momenta. This new formulation of QB has the advantage of being applicable in scattering situations. However, as we show in this brief paper, unless used with care the EF criterion of Ref.~\cite{MYD21} may fail to identify QB in the standard case of a free particle with a positive momentum.

Our paper is organized as follows. In Section~\ref{pb_sec}, we provide a concise summary of the EF formulation of QB put forward in Ref.~\cite{MYD21}. In Section~\ref{expl_ex_sec}, we analyze a concrete example of a positive-momentum state that is known to exhibit QB according to the standard criterion, Eq.~\eqref{QB_standard_crit}. We show that, for some parameter values, the EF criterion of Ref.~\cite{MYD21} can be violated for this particular state. Then, in Section~\ref{max_sec}, we address the question of the maximal backflow, computed in accordance with the EF formulation, that can be obtained for an arbitrary positive-momentum state. We demonstrate that this maximal backflow may become negligibly small in a certain parametric regime. Finally, conclusions are drawn in Section~\ref{concl_sec}.


\section{Statement of the problem}
\label{pb_sec}

A one-dimensional nonrelativistic structureless quantum particle can be described by the wave function $\psi_t(x)$ in position space, or alternatively by its Fourier transform $\widetilde{\psi}_t(p)$ in momentum space. Both functions satisfy the normalization condition
\begin{align}
\int_{\mathbb{R}} dx \, \lvert \psi_t(x) \rvert^2 = \int_{\mathbb{R}} dp \, \lvert \widetilde{\psi}_t(p) \rvert^2 = 1 \, ,
\label{normalization_wf}
\end{align}
so that $\lvert \psi_t(x) \rvert^2$ (respectively, $\lvert \widetilde{\psi}_t(p) \rvert^2$) is interpreted as the probability density of finding the particle at position $x$ (respectively, with momentum $p$) at time $t$. Hereinafter, the Fourier transform $\widetilde{g}(p)$ of a function $g(x)$ is taken to be
\begin{subequations}
\begin{align}
\widetilde{g}(p) = \frac{1}{\sqrt{2 \pi \hbar}} \int_{\mathbb{R}} dx \, e^{-ipx/\hbar} g(x) \, ,
\label{Fourier_def}
\end{align}
with the inverse transform hence given by
\begin{align}
g(x) = \frac{1}{\sqrt{2 \pi \hbar}} \int_{\mathbb{R}} dp \, e^{ixp/\hbar} \, \widetilde{g}(p) \, .
\label{inverse_Fourier_def}
\end{align}
\label{Fourier_transf_def}
\end{subequations}

As is well known (see e.g.~\cite{Lee95,Cohen}), the fact that the position and momentum observables do not commute precludes the construction of any well-defined quantum probability distribution in phase space. Instead, every quantum state can be associated with infinitely many phase-space distributions, all of them being functions of the phase-space variables $x$ and $p$, i.e., classical-like commuting variables. None of these functions however simultaneously satisfies all of the following three defining properties of probabilities, also known as Kolmogorov's axioms: positivity, normalizability, and additivity (see e.g.~\cite{Appel}). This is the reason why quantum phase-space distributions are often referred to as quasiprobability distributions. While all these distributions embed the same physical information, their mathematical properties may drastically differ from one another. Commonly used quasiprobability distributions include the Wigner, Husimi, and Glauber-Sudarshan representations.

In Ref.~\cite{MYD21}, the authors consider a particular class of phase-space distributions $f_t(x,p)$ that are everywhere positive. This property, along with the normalization condition
\begin{align}
\int_{\mathbb{R}^2} dx dp \, f_t(x,p) = 1 \, ,
\label{normalization_f_t}
\end{align}
allows one to assign a (quasi)probabilistic meaning to the distribution $f_t$. The latter is defined by
\begin{align}
f_t(x,p) = \left\lvert W_{\psi_t,\chi}(x,p) \right\rvert^2
\label{f_t_def}
\end{align}
in terms of the so-called Wigner-Moyal transform~\cite{deGosson}
\begin{multline}
W_{\psi_t,\chi}(x,p) \equiv \frac{1}{\sqrt{2\pi \hbar}} \\[0.2cm]
\times \int_{\mathbb{R}} dy \, e^{-ipy/\hbar} \chi^* \left( y - \frac{x}{2} \right) \psi_t \left( y + \frac{x}{2} \right) \, .
\label{W_chi_psi_def}
\end{multline}
The function $\chi$ represents the precision function of a measurement apparatus, and is normalized to unity\footnote{There is a slight difference between the definition of the Wigner-Moyal transform in Eq.~\eqref{W_chi_psi_def} and the way the transform is defined in Eq.~(2) of Ref.~\cite{MYD21}: The prefactor in the right-hand side of Eq.~\eqref{W_chi_psi_def} is $1/\sqrt{2 \pi \hbar}$, whereas it is $1/2 \pi \hbar$ in Ref.~\cite{MYD21}. The expression in Ref.~\cite{MYD21} is recovered from Eq.~\eqref{W_chi_psi_def} by means of the substitution $\chi(x) = \phi(x) / \sqrt{2 \pi \hbar}$.}, i.e.
\begin{align}
\int_{\mathbb{R}} dx \left\lvert \chi(x) \right\rvert^2 = 1 \, .
\label{normalization_chi}
\end{align}
It is easy to see from Eq.~\eqref{Fourier_transf_def} that $W_{\psi_t,\chi}$ can also be expressed as an integral over momentum, namely
\begin{multline}
W_{\psi_t,\chi}(x,p) = \frac{1}{\sqrt{2\pi \hbar}} \\[0.2cm]
\times \int_{\mathbb{R}} dp' \, e^{ixp'/\hbar} \, \widetilde{\chi}^* \left( p' - \frac{p}{2} \right) \widetilde{\psi}_t \left( p' + \frac{p}{2} \right) \, .
\label{W_chi_psi_momentum_repr}
\end{multline}

The ``experiment-friendly'' (EF) definition of quantum backflow (QB) put forward in Ref.~\cite{MYD21} states that QB takes place at point $x=a$ and time $t$ if the probability current $j_t(a)$ satisfies the inequality
\begin{subequations}
\begin{align}
j_t(a) < \frac{1}{m} \int_{\mathbb{R}^-} dp \, p f_t(a,p) \, .
\label{QB_criterion_MYD21_orig}
\end{align}
Hereinafter we use the notation $\mathbb{R}^{-}$ (respectively, $\mathbb{R}^{+}$) for the set of negative (respectively, positive) real numbers. Note that the criterion~\eqref{QB_criterion_MYD21_orig} can be alternatively written as
\begin{align}
J_t(a) < 0
\label{QB_criterion_MYD21_gen_current}
\end{align}
\label{QB_criterion_MYD21}
\end{subequations}
in terms of the quantity $J_t$ defined as
\begin{align}
J_t(a) \equiv j_t(a) - \frac{1}{m} \int_{\mathbb{R}^-} dp \, p f_t(a,p) \, .
\label{gen_current_def}
\end{align}
The form of condition~\eqref{QB_criterion_MYD21_gen_current} is reminiscent of the standard QB criterion~\eqref{QB_standard_crit}, with $J_t(a)$ playing the role of an effective backflow current at position $a$ and time $t$.

Let us look into the structure of criterion~\eqref{QB_criterion_MYD21} in more detail. First, note that in view of~\eqref{j_t_def} the left-hand side of Eq.~\eqref{QB_criterion_MYD21_orig} depends solely on the state $\psi_t$ of the system. However, it is clear from Eqs.~\eqref{f_t_def} and \eqref{W_chi_psi_def} that the right-hand side of Eq.~\eqref{QB_criterion_MYD21_orig} depends on $\psi_t$ as well as on the precision function $\chi$. The latter is by construction an arbitrary function, independent of the state $\psi_t$. This means that by changing the precision function $\chi$ one changes the right-hand side of Eq.~\eqref{QB_criterion_MYD21_orig} while leaving the left-hand side unchanged. Therefore, in principle, criterion~\eqref{QB_criterion_MYD21} could, for a given state $\psi_t$, be satisfied for one precision function $\chi$ but violated for another. In fact, this is precisely what we demonstrate in Section~\ref{expl_ex_sec} below.

From here on, let us focus on a Gaussian precision function of the form
\begin{subequations}
\begin{align}
\chi(x) = \frac{1}{\pi^{1/4} \sqrt{\sigma}} \, e^{-x^2/2 \sigma^2} \,.
\label{Gaussian_chi_pos}
\end{align}
The corresponding momentum representation is given by
\begin{align}
\widetilde{\chi}(p) = \frac{1}{\pi^{1/4} \sqrt{\widetilde{\sigma}}} \, e^{-p^2/2 \widetilde{\sigma}^2} \,.
\label{Gaussian_chi_mom}
\end{align}
\label{Gaussian_chi}%
\end{subequations}
Here, the position- and momentum-space widths $\sigma$ and $\widetilde{\sigma}$, respectively, are related via $\sigma \widetilde{\sigma} = \hbar$. Our motivation for this choice of $\chi$ is twofold. First, a Gaussian smoothing function is the most natural choice for mimicking a finite precision of a measurement apparatus\footnote{For instance, measurements are modeled by Gaussian quasiprojectors in~\cite{YHH12}.}, and, as such, is the precision function used in all examples of Ref.~\cite{MYD21}. Second, it follows from Eqs.~\eqref{f_t_def} and \eqref{W_chi_psi_def} that the phase-space distribution $f_t$ obtained using a Gaussian $\chi$ corresponds to the Husimi distribution (see e.g. Eq. (7.25) in~\cite{Lee95}), which is arguably the most commonly used nonnegative quantum phase-space distribution function.

Our aim is to compare the EF criterion~\eqref{QB_criterion_MYD21} against the standard definition of QB, based on Eq.~\eqref{QB_standard_crit}. The latter only applies to a free particle in a positive-momentum state, i.e., a state described by the momentum wave function
\begin{align}
	\widetilde{\psi}_t(p) = e^{-ip^2t/2m\hbar} \widetilde{\psi}_0(p)
	\label{time_dep_mom_wf}
\end{align}
that vanishes identically for negative momenta,
\begin{align}
	\widetilde{\psi}_t(p) = 0 \qquad \text{if} \qquad p < 0 \, .
	\label{positive_mom_wf}
\end{align}
So, below we only consider wave functions of this form.

In the present paper, we argue that the usefulness of the EF definition of QB, given by Eq.~\eqref{QB_criterion_MYD21}, is strongly dependent on the width $\widetilde{\sigma}$ of the precision function. We present our argument in the following two sections. Thus, in Section~\ref{expl_ex_sec}, we consider the case of a particle prepared in a particular positive-momentum state $\widetilde{\psi}_0$ that satisfies the standard criterion~\eqref{QB_standard_crit} of QB, and show that the EF criterion~\eqref{QB_criterion_MYD21} gets violated if $\widetilde{\sigma}$ exceeds a certain value. Then, in Section~\ref{max_sec}, we address the problem of the maximal backflow probability transfer $\Delta_{\max}$ achievable with positive-momentum states. On the one hand, numerical analysis based on the standard backflow criterion~\eqref{QB_standard_crit} yields~\cite{BM94,EFV05,PGK06}
\begin{align}
\Delta_{\text{max}}^{(\text{BM})} \approx 0.0384517 \, ,
\label{BM_constant}
\end{align}
a number commonly referred to as the Bracken-Melloy bound. On the other hand, we demonstrate that $\Delta_{\max}$ inferred from the EF criterion~\eqref{QB_criterion_MYD21} decays with $\widetilde{\sigma}$ and can become arbitrarily small. This means that the new criterion~\eqref{QB_criterion_MYD21} may fail to identify QB if the precision function is too broad in momentum space. We further quantify the range of $\widetilde{\sigma}$ in which the EF definition of QB is consistent with the standard one.


\section{Explicit example}
\label{expl_ex_sec}

Let us consider the particular positive-momentum state that was introduced in Ref.~\cite{BM94} as a concrete example of QB for normalized states. The state corresponds to a fixed instant in time, taken to be $t=0$, and is given by
\begin{subequations}
\begin{equation}
\widetilde{\psi}_0(p) \equiv \left\{\begin{array}{ll}
0  & , \quad p < 0 \\[0.3cm]
\frac{18 p}{\sqrt{35 \alpha^3}} \left( e^{-p/\alpha} - \frac{1}{6} \, e^{-p/2\alpha} \right) & , \quad p > 0
\end{array}\right.
\label{psi_0_mom_expr}
\end{equation}
where $\alpha$ is a positive constant that has the dimension of a momentum. Note that the function $\widetilde{\psi}_0(p)$ is continuous at $p=0$. Substituting Eq.~\eqref{psi_0_mom_expr} into Eq.~\eqref{inverse_Fourier_def}, one obtains the corresponding position representation~\cite{BM94}
\begin{align}
\psi_0(x) = 18 \sqrt{\frac{\alpha \hbar^3}{70 \pi}} \left[ \frac{1}{(\hbar - i \alpha x)^2} - \frac{2}{3} \frac{1}{(\hbar - 2 i \alpha x)^2} \right] \, .
\label{psi_0_pos_expr}
\end{align}
\label{expl_ex_wf}
\end{subequations}
The probability current $j_0(0)$ at $a=0$, calculated according to Eq.~\eqref{j_t_def}, yields~\cite{BM94}
\begin{align}
j_0(0) = - \frac{36 \alpha^2}{35 \pi m \hbar} < 0 \, .
\label{init_current_expr}
\end{align}
The criterion~\eqref{QB_standard_crit} is thus clearly satisfied, meaning that the particular positive-momentum state~\eqref{expl_ex_wf} is a backflowing state in the standard sense.

We now investigate the predictions of the EF criterion~\eqref{QB_criterion_MYD21} for the state~\eqref{expl_ex_wf}. To this end, we first compute the corresponding phase-space distribution $f_0(0,p)$ obtained from Eq.~\eqref{f_t_def} for $x=0$ and $t=0$. In order to focus on the role played by the width $\widetilde{\sigma}$ of the Gaussian precision function, we use a system of units that is adapted to this particular state by introducing the dimensionless momentum
\begin{align}
\eta \equiv \frac{p}{\alpha} \, .
\label{dimensionless_mom}
\end{align}
Then, combining the definitions~\eqref{f_t_def} and~\eqref{W_chi_psi_momentum_repr} with Eqs.~\eqref{Gaussian_chi_mom},~\eqref{psi_0_mom_expr} and~\eqref{dimensionless_mom}, we find
\begin{align}
f_0 (0,\alpha\eta) = \frac{162}{35 \pi^{3/2} \hbar} \frac{I^2 (\eta;s)}{s} \, ,
\label{f_0_expr}
\end{align}
where $s$ is the dimensionless momentum width of the precision function,
\begin{align}
s \equiv \frac{\widetilde{\sigma}}{\alpha} \, ,
\label{width_dimensionless}
\end{align}
and $I$ is given by the integral
\begin{align}
I (\eta;s) \equiv \int_{\mathbb{R}^+} d\eta' \, \eta' e^{- (\eta' - \eta)^2/2 s^2} \left( e^{-\eta'} - \frac{e^{-\eta'/2}}{6} \,  \right) .
\label{I_def}
\end{align}
Alternatively, $I$ can be expressed as
\begin{multline}
I(\eta;s) = s e^{-\eta^2/2s^2} \left\{ \frac{5s}{6} \right. \\
\left. - \sqrt{\frac{\pi}{2}} (s^2-\eta) e^{(s^2-\eta)^2/2s^2} \operatorname{erfc} \left( \frac{s^2-\eta}{\sqrt{2}s} \right) \right. \\[0.2cm]
\left. + \sqrt{\frac{\pi}{2}} \frac{s^2-2\eta}{12} e^{(s^2-2\eta)^2/8s^2} \operatorname{erfc} \left( \frac{s^2-2\eta}{2\sqrt{2}s} \right) \right\}
\label{I_expr}
\end{multline}
with
\begin{align}
\erfc(z) = 1 - \erf(z) = \frac{2}{\sqrt{\pi}} \int_{z}^{\infty} dy \, \mathrm{e}^{-y^2}
\label{erfc_def}
\end{align}
being the complementary error function.

Substituting Eqs.~\eqref{init_current_expr} and \eqref{f_0_expr} into Eq.~\eqref{gen_current_def}, we obtain the effective backflow current:
\begin{align}
J_0(0) = - \frac{18}{35 \pi} \frac{\alpha^2}{m \hbar} \left[ 2 + \frac{9}{\sqrt{\pi} s} \int_{\mathbb{R}^-} d\eta \, \eta I^2 (\eta;s) \right] \, .
\label{J_0_0_expr}
\end{align}
It is clear from this expression that the sign of $J_0(0)$ only depends on the (dimensionless) width $s$ of the precision function. In particular, note that the integral in the right-hand side of~\eqref{J_0_0_expr} is necessarily negative.


\begin{figure}[ht]
\centering
\includegraphics[width=0.48\textwidth]{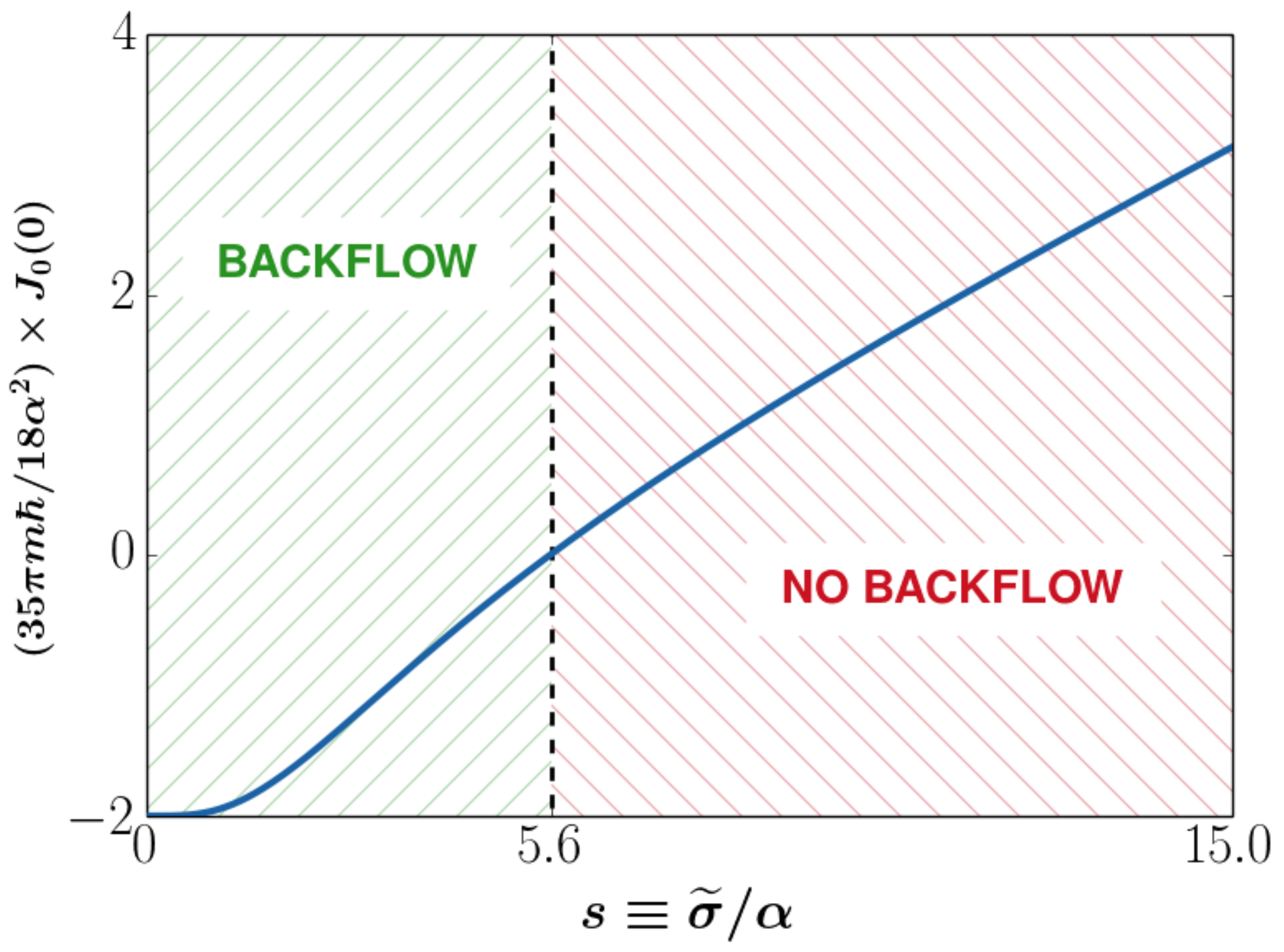}
\caption{Scaled effective backflow current (solid blue curve), as given by~\eqref{J_0_0_expr}, as a function of the (dimensionless) width $s$ of the Gaussian precision function.}
\label{expl_ex_fig}
\end{figure}


We now evaluate $J_0(0)$ for various values of $s$ by numerically computing the integral in Eq.~\eqref{J_0_0_expr}. The results are shown in Fig.~\ref{expl_ex_fig}, where the (scaled) effective backflow current is depicted by the solid blue curve. We see that for small enough values of $s$ the current $J_0(0)$ takes negative values (hashed green region). In this regime, the EF criterion~\eqref{QB_criterion_MYD21_gen_current} predicts the occurrence of QB and is in agreement with the standard backflow criterion~\eqref{QB_standard_crit}. However, for $s \gtrsim 5.6$ the current $J_0(0)$ takes positive values (hashed red region). In this regime, condition~\eqref{QB_criterion_MYD21} is no longer fulfilled, implying that the EF criterion fails to identify QB.

This shows that the compatibility between the EF criterion~\eqref{QB_criterion_MYD21} and the standard one, Eq.~\eqref{QB_standard_crit}, is sensitive to the momentum width $\widetilde{\sigma}$ of the precision function. While this conclusion was reached based on the particular state~\eqref{expl_ex_wf}, we now investigate the predictions of condition~\eqref{QB_criterion_MYD21} for general positive-momentum states.


\section{Maximal backflow}
\label{max_sec}

We now consider the following question, originally posed in Ref.~\cite{BM94}: For a free particle in a positive-momentum state and a given time interval $(0,T)$, what is the maximal amount of probability $\Delta_{\max}$ that can possibly cross the space point $x=0$ in the ``wrong'' direction? If one relies on the standard definition of QB, Eq.~\eqref{QB_standard_crit}, the answer to this question is given by the Bracken-Melloy bound, Eq.~\eqref{BM_constant}. Here, we want to answer this question adopting the new, EF definition of QB, Eq.~\eqref{QB_criterion_MYD21}, and compare the results with the standard case.

To this end, we use the approach developed in Ref.~\cite{BM94}. The particle is described by the wave function
\begin{align}
\psi_t(x) = \frac{1}{\sqrt{2 \pi \hbar}} \int_{\mathbb{R}^+} dp \, e^{ixp/\hbar} \, e^{-ip^2t/2m\hbar} \widetilde{\psi}_0(p) \, .
\label{pos_wf_Fourier_expr}
\end{align}
Substituting Eq.~\eqref{pos_wf_Fourier_expr} into Eq.~\eqref{j_t_def} and setting $a=0$, we obtain
\begin{multline}
j_t(0) = \frac{1}{4 \pi m \hbar} \int_{\left( \mathbb{R}^+ \right)^2} dp dp' \, \widetilde{\psi}_0^*(p) \\
\times (p+p') e^{it \left( p^2 - p'^2 \right)/2m\hbar} \widetilde{\psi}_0(p') \, .
\label{j_t_0_expr}
\end{multline}
Then, upon combining the definitions~\eqref{f_t_def} and~\eqref{W_chi_psi_momentum_repr} with Eqs.~\eqref{Gaussian_chi_mom},~\eqref{time_dep_mom_wf}, and~\eqref{positive_mom_wf}, we get
\begin{multline}
\frac{1}{m} \int_{\mathbb{R}^-} dp \, p f_t(0,p) = \frac{1}{4 \pi m \hbar} \int_{\left( \mathbb{R}^+ \right)^2} dp dp' \, \widetilde{\psi}_0^*(p) \\
\times U(p,p';\widetilde{\sigma}) e^{it \left( p^2 - p'^2 \right)/2m\hbar} \widetilde{\psi}_0(p') \, ,
\label{neg_p_int_expr}
\end{multline}
where the function $U$ is defined as
\begin{align}
&U(p,p';\widetilde{\sigma}) \nonumber \\ & \quad \equiv \frac{2}{\sqrt{\pi} \widetilde{\sigma}} \int_{\mathbb{R}^-} dp'' \, p'' e^{- [ (p''-p)^2 + (p''-p')^2] / 2 \widetilde{\sigma}^2} \,.
\label{U_def}
\end{align}
The function $U$ can also be expressed in terms of the complementary error function (see Eq.~\eqref{U_dimensionless_def} below). Substituting Eqs.~\eqref{j_t_0_expr} and~\eqref{neg_p_int_expr} into Eq.~\eqref{gen_current_def} yields the effective backflow current
\begin{multline}
J_t(0) = \frac{1}{4 \pi m \hbar} \int_{\left( \mathbb{R}^+ \right)^2} dp dp' \, \widetilde{\psi}_0^*(p) \\
\times \left[ p+p' - U(p,p';\widetilde{\sigma})\right] e^{it \left( p^2 - p'^2 \right)/2m\hbar} \widetilde{\psi}_0(p') \, .
\label{gen_current_expr}
\end{multline}

The backflow probability transfer through $x=0$ over the time interval $0 < t < T$ is given by
\begin{align}
\Delta \equiv - \int_{0}^{T} dt \, J_t(0) \,.
\label{Delta_def}
\end{align}
The maximization of $\Delta$ is performed under the constraint that $\widetilde{\psi}_0$ is normalized according to Eq.~\eqref{normalization_wf}. Substituting Eq.~\eqref{gen_current_expr} into Eq.~\eqref{Delta_def}, evaluating the time integral, and using the method of Lagrange multipliers, we find that the maximum of $\Delta$ (subject to the normalization constraint) is given by the largest eigenvalue $\Delta_{\text{max}}$ $(\equiv \sup\limits \lambda)$ of the following integral eigenvalue problem:
\begin{multline}
\int_{\mathbb{R}^+} dp' \, \frac{i}{2 \pi} \frac{p+p' - U(p,p';\widetilde{\sigma})}{p^2-p'^2} \\
\quad \times \left[ e^{iT \left( p^2 - p'^2 \right)/2m\hbar} - 1 \right] \widetilde{\psi}_0(p') = \lambda \widetilde{\psi}_0(p) \, .
\label{eigenval_eq}
\end{multline}

We now rewrite Eq.~\eqref{eigenval_eq} in a dimensionless form. Making the change of variables
\begin{align}
u \equiv \sqrt{\frac{T}{4 m \hbar}} \, p \qquad \text{and} \qquad u' \equiv \sqrt{\frac{T}{4 m \hbar}} \, p' \, ,
\label{u_def}
\end{align}
defining the dimensionless width $\varsigma$ of the precision function as
\begin{align}
\varsigma \equiv \sqrt{\frac{T}{m \hbar}} \, \widetilde{\sigma} \, ,
\label{width_dimensionless_max}
\end{align}
and introducing the dimensionless functions
\begin{align}
&\mathcal{U}(u,u';\varsigma) \nonumber \\ &\quad \equiv \sqrt{\frac{T}{4 m \hbar}} \, U\left( \sqrt{\frac{4 m \hbar}{T}} u, \sqrt{\frac{4 m \hbar}{T}} u';\sqrt{\frac{m \hbar}{T}} \varsigma \right) \nonumber \\[0.1cm]
&\quad = \frac{u+u'}{2} e^{-(u-u')^2/\varsigma^2} \operatorname{erfc} \left( \frac{u+u'}{\varsigma} \right) \nonumber \\ &\qquad\qquad - \frac{\varsigma e^{-2(u^2 + u'^2)/\varsigma^2}}{2 \sqrt{\pi}}
\label{U_dimensionless_def}
\end{align}
and
\begin{align}
\varphi(u) \equiv \left( \frac{4 m \hbar}{T} \right)^{1/4} e^{-iu^2} \widetilde{\psi}_0 \left( \sqrt{\frac{4 m \hbar}{T}} u \right) \, ,
\label{varphi_def}
\end{align}
we arrive at the following dimensionless integral eigenproblem:
\begin{multline}
\int_{\mathbb{R}^+} du' \, \frac{u+u' - \mathcal{U}(u,u';\varsigma)}{\pi(u'^2-u^2)} \sin (u^2 - u'^2) \varphi(u') \\ = \lambda \varphi(u) \, .
\label{eigenval_eq_dimensionless}
\end{multline}


\begin{figure}[ht]
\centering
\includegraphics[width=0.48\textwidth]{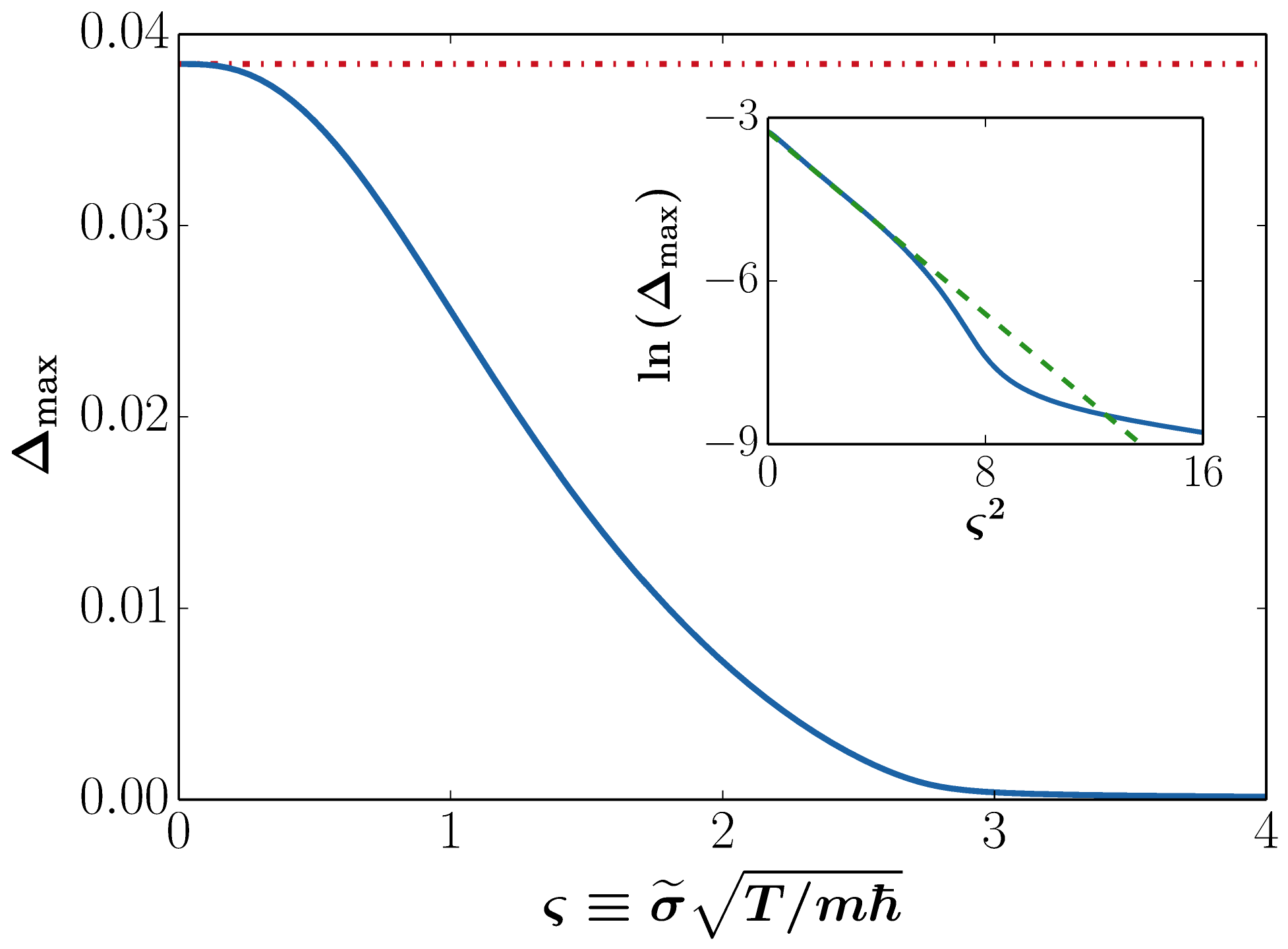}
\caption{Behavior of the maximal eigenvalue $\Delta_{\text{max}}$ (solid blue curve) of the eigenvalue equation~\eqref{eigenval_eq_dimensionless} with respect to the (dimensionless) width $\varsigma$ of the Gaussian precision function. The dash-dotted horizontal red line represents the Bracken-Melloy bound~\eqref{BM_constant}. Inset: Logarithm of $\Delta_{\text{max}}$ as a function of $\varsigma^2$ (solid blue curve). The dashed green curve shows a Gaussian decay.}
\label{max_backflow_fig}
\end{figure}


We then numerically solve the eigenvalue equation~\eqref{eigenval_eq_dimensionless} and evaluate the largest eigenvalue $\Delta_{\text{max}}$\footnote{The numerical method is based on the approach originally presented in Ref.~\cite{BM94}, and has been further discussed in Section IIIB of Ref.~\cite{Gou20_1} and the appendix of Ref.~\cite{Gou20_2}.}, i.e. the maximal backflow, for different values of the (dimensionless) width $\varsigma$. The results are presented in Fig.~\ref{max_backflow_fig}. The solid blue curve shows the behavior of $\Delta_{\text{max}}$ as a function of $\varsigma$, and the dash-dotted horizontal red line represents the Bracken-Melloy bound, Eq.~\eqref{BM_constant}.

Our first observation is that $\Delta_{\max}$ approaches the Bracken-Melloy bound as $\varsigma \to 0$. This means that for small enough widths, i.e., for a Gaussian precision function that is sufficiently narrow in momentum space (and, consequently, sufficiently broad in position space), the EF criterion~\eqref{QB_criterion_MYD21} allows for the same maximal backflow probability transfer as the standard criterion~\eqref{QB_standard_crit}. This fully agrees with the fact, explicitly noted in Ref.~\cite{MYD21}, that in the limiting case of a precision function given by a Dirac $\delta$-function in momentum space, the EF formulation~\eqref{QB_criterion_MYD21} of QB reduces to the standard one, Eq.~\eqref{QB_standard_crit}, for positive-momentum states.

Our second observation is that $\Delta_{\text{max}}$ monotonically decreases with $\varsigma$. This means that the EF criterion~\eqref{QB_criterion_MYD21} is less efficient at signalling the presence of QB at larger momentum widths of the precision function. Furthermore, Fig.~\ref{max_backflow_fig} strongly suggests that $\Delta_{\text{max}}$ vanishes in the limit $\varsigma \to \infty$.

At first sight, it might seem that the curve $\Delta_{\text{max}} (\varsigma)$ follows a Gaussian decay. This is however not the case, as is clear from the inset in Fig.~\ref{max_backflow_fig}. The latter indeed shows that the logarithm of $\Delta_{\text{max}}$ (solid blue curve) depends on $\varsigma^2$ in a more intricate way than a simple linear dependence (dashed green curve).

Figure~\ref{max_backflow_fig} allows us to estimate the parametric regime in which the EF criterion~\eqref{QB_criterion_MYD21} is capable of identifying QB. At the practical level, the backflow probability transfer is only appreciable for $\varsigma \lesssim 1$. (We can see from the data that $\Delta_{\text{max}}=\frac{1}{2} \Delta_{\text{max}}^{(\text{BM})}$ for $\varsigma \approx 1.29$.) This condition can, in view of Eq.~\eqref{width_dimensionless_max}, be interpreted as follows: A measurement apparatus, characterized by a momentum precision $\widetilde{\sigma}$, may be able to detect QB only if the measurement is performed on a time scale $T$ satisfying 
\begin{align}
T \lesssim \frac{m \hbar}{\widetilde{\sigma}^2} \, .
\label{time_scale}
\end{align}


\section{Conclusion}
\label{concl_sec}

We have investigated the compatibility of two formulations of quantum backflow (QB) -- on the one hand, the original formulation due to Bracken and Melloy~\cite{BM94}, with the backflow criterion given by Eq.~\eqref{QB_standard_crit}, and, on the other hand, the ``experiment-friendly'' (EF) formulation, with criterion~\eqref{QB_criterion_MYD21}, recently proposed in Ref.~\cite{MYD21}. In order to make a direct comparison, we applied both formulations to the case of a free particle in a positive-momentum state. The EF criterion, Eq.~\eqref{QB_criterion_MYD21}, involves a free parameter $\widetilde{\sigma}$, playing the role of a momentum-space width of the precision function of a measurement apparatus. Our main conclusion, in a nutshell, is that the two formulations of QB are compatible only if $\widetilde{\sigma}$ is substantially small, i.e., only if the measurement apparatus is sufficiently precise in momentum space.

More specifically, we have considered a concrete example of a normalized positive-momentum state, given by Eq.~\eqref{expl_ex_wf}, that exhibits QB in the standard sense of criterion~~\eqref{QB_standard_crit}. We have demonstrated the existence of a critical value of the width $\widetilde{\sigma}$ above which the EF criterion~\eqref{QB_criterion_MYD21} is not satisfied, implying that, in the latter parameter range, the EF formulation of QB is in conflict with the standard one.

Then, in the context of the EF formulation of QB, we have investigated the maximal backflow probability transfer $\Delta_{\max}$ through a spatial point over a fixed time interval. We have numerically determined $\Delta_{\text{max}}$ as a function of the width $\widetilde{\sigma}$. Our analysis indicates the following behavior of this function. As $\widetilde{\sigma}$ tends to zero, $\Delta_{\max}$ approaches the Bracken-Melloy bound, given by Eq.~\eqref{BM_constant}. Then, as $\widetilde{\sigma}$ increases, the maximal backflow transfer decays monotonously and vanishes in the limit $\widetilde{\sigma} \to \infty$. This allows us to identify the following parametric regime in which the EF criterion~\eqref{QB_criterion_MYD21} is capable of signaling the occurrence of QB: The measurement apparatus has to operate on a sufficiently short time scale $T$, satisfying condition~\eqref{time_scale}, in order to make the detection of QB practically feasible.

The EF formulation aims at providing a definition of QB that is better suited to an experimental investigation as compared to the standard one. Therefore, a brief discussion of the possibility of an experimental validation of the EF backflow is in order. Checking the validity of Eq.~\eqref{QB_criterion_MYD21_orig} experimentally requires to measure the probability current $j_t(a)$ and, in addition, to evaluate the momentum integral of $p f_t(a,p)$. As pointed out e.g. in Ref.~\cite{BM94}, the problem of measuring $j_t(a)$ can be mapped onto a more conventional problem of measuring an electric current if one conducts the experiment with electrically charged particles.\footnote{The relation between the electric and probability currents is discussed in section 2.4 of Ref.~\cite{Gottfried}.} Evaluating the momentum integral of $p f_t(a,p)$ is clearly a more challenging problem, as it requires one's ability to experimentally access the phase-space distribution $f_t$, i.e.~the Husimi function in our case. This challenge however is not beyond the reach of cutting-edge experimental techniques. As stated in Appendix A of Ref.~\cite{HarocheRaimond}, the Husimi function is the expectation value of an observable and, as such, is a measurable quantity. While more ingenious strategies may exist, an experimental measurement of the wave function itself (using, e.g., techniques developed in Refs.~\cite{LSP11,PXK19,ZZM19,SCP20}) can in principle be used to reconstruct the Husimi distribution.

One of the main appeals of the EF formulation of QB, introduced in Ref.~\cite{MYD21}, is that it offers a promising approach for studying QB in situations where a particle moves in the presence of external forces and in systems of many interacting particles. However, as we have shown in this paper, care must be exercised when using the EF criterion~\eqref{QB_criterion_MYD21}: One must make sure that the measurement apparatus is sufficiently precise in momentum space and operates on short enough time scales.


\printbibliography


\end{document}